\documentclass[11pt, conference]{IEEEtran}
\usepackage{latexsym}
\usepackage{amsmath}
\usepackage{amssymb}
\usepackage{amsfonts}
\usepackage{graphicx}
\usepackage{pstricks}
\usepackage{pst-node,pst-plot}
\usepackage{setspace}
\newtheorem{corollary}{Corollary}
\newtheorem{theorem}{Theorem}

\newcommand{\m}[1]{\ensuremath{\mathbf{#1}^n}}
\title{What is needed to exploit knowledge of primary transmissions?}
\author{\authorblockN{Pulkit Grover}
\authorblockA{Wireless Foundations, EECS Department \\
University of California at Berkeley\\
Email: pulkit\;@\;eecs.berkeley.edu}
\and
\authorblockN{Anant Sahai}
\authorblockA{Wireless Foundations, EECS Department\\
University of California at Berkeley\\
Email: sahai\;@\;eecs.berkeley.edu}}

\begin{document}
\maketitle
\begin{abstract}
  Recently, Tarokh and others have raised the possibility that a
  cognitive radio might know the interference signal being transmitted
  by a strong primary user in a non-causal way, and use this knowledge
  to increase its data rates. However, there is a subtle difference
  between knowing the signal transmitted by the primary and the actual
  interference at our receiver since there is a wireless channel
  between these two points. We show that even an unknown phase results
  in a substantial decrease in the data rates that can be achieved,
  and thus there is a need to feedback interference channel estimates
  to the cognitive transmitter. We then consider the case of fading channels. We derive an upper bound on the rate for given outage error probability for faded dirt. We give a scheme that uses appropriate
  ``training" to obtain such estimates and quantify this scheme's
  required overhead as a function of the relevant coherence time and
  interference power.
\end{abstract}
\section{Introduction}

Knowledge of different aspects of a wireless channel can be exploited
in various ways to increase the achievable data rates. For example,
consider an OFDM transmission when the channel has frequency selective
fading. The receiver needs to know the channel in order to do
equalization and thereby interpret the information on the various
subcarriers. To support this equalization, the transmitter helps by
dedicating some of its energy to transmitting known pilot tones
\cite{davidbook}. As the bandwidth gets large and the number of fading
parameters increases, the overhead required to learn these parameters
at the receiver increases and it has been shown \cite{telatar, zheng}
that the optimal signaling in wideband communications is `peaky' ---
concentrating most of its energy in a few time/frequency slots. 

The receiver clearly needs to implicitly or explicitly learn the
wireless channel, but in many cases, the system can benefit from
having the transmitter exploit this knowledge as well. If the channel
is frequency selective, then the transmitter can use the simple
water-pouring scheme \cite{davidbook} to allocate its power across
subchannels to achieve optimal rates. \cite{verduconstellations}
extends this insight to practical settings where the signal
constellations are constrained. If the fading is not frequency
selective, then for a single-input single-output channel, transmitter
knowledge does not really impact waveform design.

In practice, channel knowledge is never perfect. Channel uncertainty
at the receiver comes from the finite underlying coherence time of the
time-varying wireless channel. In wireless systems that time-share a
single frequency between forward and backward links, wireless
reciprocity induces a similar channel uncertainty at the transmitter.
More typically, the limitation in channel knowledge at the transmitter
comes from the quality of the feedback link over which the channel
information is sent back.  Such uncertainty in knowledge of the
channel is known to result in appreciable degradation in
performance\cite{wigger, lapidothshamai, jindal}.  Even for flat
fading, \cite{wigger} shows that for the MIMO broadcast channel,
fading uncertainty reduces the throughput substantially at high SNR.

This paper explores the impact of channel uncertainty in the cognitive
radio context. Generally, a cognitive radio senses its environment,
learns from it, and opportunistically uses the channel resources.
While in many cases, the focus is on finding empty bands in which to
transmit, \cite{tarokh,viswanath,wu} suggest a radically different
perspective. They suggest that very strong interference from a primary
user can be exploited by a communication system to increase its
achievable rate beyond what is possible by simply treating the
interference as noise.

If the receiver perfectly knows the interfering signal, it can simply
subtract it off from the received signal and thereby pretend that
there is no interference. It makes no difference if the knowledge is
delayed, instantaneous, or non-causal. The idea that noncausal
knowledge could be useful at the {\em transmitter} is attributed to
Gelfand and Pinsker \cite{gelfand}. In a surprising result, Costa
\cite{costa} proved that the capacity of a channel with additive
Gaussian interference (known non-causally at the transmitter) is same
as that with no interference at all. This is called ``dirty paper
coding'' and it forms the basis for the schemes in
\cite{tarokh,viswanath,wu} where it is assumed that the transmitter is
able to decode what the primary is sending before it begins its own
transmissions.

\begin{figure}
\includegraphics[width=3in]{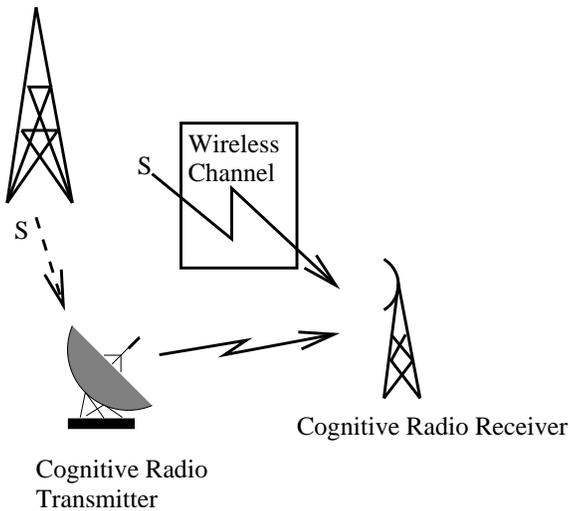}
\caption{A cognitive radio cognizant of signal transmitter by the primary. For simplicity, we assume that the interferer's channel results in a phase shift, $e^{j\theta}$.}
\label{fig:cognitiveradio}
\end{figure}

Fig.~\ref{fig:cognitiveradio} illustrates the setup. Once again, it is
natural to consider uncertainty in the knowledge of fading. After all,
there is a difference between knowing the transmitted primary signal
and knowing the interference caused by this signal at our own
receiver. We consider the case of flat fading and concentrate on the
unknown phase shift for simplicity. We pose the following questions:
\begin{itemize}
 \item How important is it for the secondary {\em transmitter} to have precise
interference phase knowledge? 
 \item How does the fade amplitude affect the achievable rates? 
 \item Does feedback help and if so, how should we use it? 
\end{itemize}
 
The problem of uncertainty in knowledge of the interference at the
transmitter has been addressed in \cite{khisti}\cite{mitran}.
\cite{khisti} models interference uncertainty by allowing the channel
interference vector to take one of two possible values, both known at
the transmitter only. The two possible interference vectors are drawn
independently with Gaussian distributions. The transmitter does not
know which particular value the interference assumes. \cite{khisti}
shows that situation is equivalent to a broadcast channel model with a
common message, where the transmitter knows the interference vectors
to the two receivers. They are able to show an explicit rate penalty
for not knowing the interference exactly. For the scenario in
Fig.~\ref{fig:cognitiveradio}, however, the model in \cite{khisti} is
not realistic: the possible realizations of the interference vectors
are far from independent. It is unclear from \cite{khisti} whether
phase-uncertainty alone would impose a significant rate penalty.

\cite{mitran} considers the general problem of uncertainty in
knowledge of interference, modeling it as a compound Gelfand-Pinsker
problem. This model does capture the scenario in
Fig.~\ref{fig:cognitiveradio} and gives the following lower bound $C_l$
and upper bound $C_u$ on the capacity $C$.
\begin{equation} \label{eqn:mitranlower}
C_l=\sup_{P_{U|X,S,W},P_{X|S,W},P_W}\inf_{\beta\in\mathcal{C}}[I^{
\beta}(U;Y|W)-I(U;S|W))]
\end{equation}
\begin{equation} \label{eqn:mitranupper}
C_u=\sup_{P_{X|S,W},P_W}\inf_{\beta\in\mathcal{C}}[\sup_{P_{U|X,S,W}}I^{
\beta}(U;Y|W)-I(U;S|W))]
\end{equation}
where $\beta\in\mathcal{C}$ is a parameter characterizing the uncertainty in the knowledge of state $S$.
Notice that both bounds are expressed as supremums over the
distribution of an auxiliary random variable $U$. Evaluated for any
particular $U$, (\ref{eqn:mitranlower}) gives a lower bound to the
capacity. However, any particular $U$ {\em does not} give a valid
upper bound in (\ref{eqn:mitranupper}). That requires taking a
supremum over all possible $U$'s, and it is not clear from
\cite{mitran} what the appropriate choice of $U$ is. Without such
guidance, the upper bound (\ref{eqn:mitranupper}) is not computable
for even a simple wireless channel model.

In addition to the bounds, a scheme is suggested in \cite{mitran} for
flat fading channels. The scheme uses Costa's dirty-paper strategy,
where the auxiliary random variable $U$ is defined to be equal to
$X+\alpha S$.  A seemingly reasonable value of $\alpha$ is chosen,
according to the distributions of the fade parameters. Unfortunately,
the constant $\alpha$ becomes zero if the fading coefficient has zero
mean. Since Rician fading has a non-zero mean, \cite{mitran}
concentrates on that case. For simple phase
uncertainty, the strategy is not optimal since $\alpha=0$ implies
ignoring the interference knowledge at the transmitter. In
Section~\ref{sec:sector} we show a scheme (inspired by \cite{khisti})
that performs better at low SINR.

Sacrificing generality, we obtain stronger bounds than \cite{mitran}
for particular case of phase uncertainty. Section~\ref{sec:bound}
shows that the lack of phase knowledge can substantially reduce the attainable 
rates. This is done by modifying the bounding strategy in
\cite{khisti}. Section~\ref{sec:feedback} considers the case when
low-rate feedback is available from the receiver. We show that if the
transmitter encodes certain ``training data'', an asymptotically
vanishing rate-loss is incurred by having to learn the phase to some
fidelity and and communicate it back to the transmitter. For the
residual uncertainty in phase, we give a modified dirty-paper coding
scheme that achieves the capacity for perfect phase knowledge in the
limit of uncertainty going to zero.

We then use the de consider the problem of fading uncertainty regarding the primary's transmission, where the magnitude of the fade is also uncertain. In the classical point-to-point communication problem, a deep channel fade necessarily introduces errors, even with complete knowledge of the fading coefficient. Therefore, the problem of interest there is finding the achievable rates for given outage probability. However, here we consider fading only for the primary's signal, which is really the interference to the secondary. With complete knowledge of the fading coefficient, the problem reduces to Costa's dirty-paper coding problem~\cite{costa} and there is no channel outage due to known interference fading. 

Consider Rayleigh fading of the primary's signal. Suppose our scheme ignores the knowledge of the primary's transmission. The magnitude of the fading coefficient of the primary's transmission can be arbitrarily large. Hence, the interference power can be arbitrarily large too. The SINR, and hence rate, achieved by this scheme would therefore be zero! This suggests considering achievable rates for a given outage probability. On the other hand, the trivial upper bound, corresponding to complete knowledge of the fading coefficient, suggests that outage probability does not change the rate significantly. Therefore, it is not clear if allowing for some outage probability would make a significant difference to the achievable rate. 

From a system design perspective, we conclude that in designing
systems that exploit non-causal knowledge of the interference,
feedback has an important role to play. Furthermore, the natural
``broadcast nature'' of wireless transmissions will be destroyed since
the transmitted data is targeted at a particular phase of the
interference. This means that schemes like
\cite{wirelessnetworkcoding} will not function in this environment.

\section{Problem Statement}
\label{sec:prob}
Consider the following complex channel model.
\begin{equation}
Y_i=X_i+h_iS_i+Z_i
\end{equation}
where $Y_i$ denotes the received signal at the secondary receiver at
the time instant $i$, $X_i$ denotes the signal transmitted by the
secondary transmitter, $S_i$ denotes the signal of the primary, $h_i$
is the channel fade coefficient, and $Z_i$ is AWGN. We are interested
in maximizing the rate for the secondary.

For simplicity, we first assume that the fade amplitude, $|h_i|=1$, and the
phase of $h_i$ is constant $\theta$, independent of $i$. Also, we
assume block encoding and decoding, with block-length $n$. Therefore,
the channel model is
\begin{equation}
\m{Y}=\m{X}+\m{S}e^{j\theta}+\m{Z}
\end{equation}
where the superscript $n$ denotes an $n-$length vector. The
interference is modeled as (complex) Gaussian i.i.d at each instant, with
variance $Q$ and mean zero. The signal $X$ is power constrained, in that
the average power should not exceed $P$. The noise is AWGN, and
without loss of generality, its variance is assumed to be 1. Under
this model, we seek to find the maximum achievable rate under no
knowledge of the phase $\theta$ at the secondary transmitter.

We then consider the case when the fade amplitude $|h_i|=\gamma$ and the phase in again unknown. The channel model now is
\begin{equation}
\m{Y}=\m{X}+\m{S}e^{j\theta}+\m{Z}
\end{equation}
We want to find the achievable rate region for unknown $\gamma$ and $\theta$.

In what follows, $W$ denotes the message to be communicated by the
cognitive radio. $W$ is assumed to be chosen uniformly from the set of
$2^{nR}$ messages.

\section{Upper bound on the communication rate under phase uncertainty of the interference}
\label{sec:bound2}
In this section, we give an upper bound on the communication rate
assuming the phase of the interference is unknown at the transmitter.

In \cite{khisti}, the authors found bounds on the rate assuming that
interference vector lies in a two-point set known to the transmitter.
They observed that the problem of unknown interference is the same as
that for communicating the same message simultaneously to two
receivers, where the receivers face different interference vectors. We
build on their idea.

Consider the case when the transmitter is unsure whether the phase at
the receiver is $0$ or $\phi$. This uncertainty is certainly no worse
than no knowledge of the phase. Hence, any upper bound for this
two-point uncertainty set is an upper bound for our case of no phase
knowledge.

Suppose the uncertainty is between $\m{S}$ and $\m{S}e^{j\phi}$, for
some fixed $\phi\leq \pi$. Denote by $\m{Y}_1$ the output
corresponding to the first user and $\m{Y}_2$ as the output
corresponding to the second user. Then, 
\begin{eqnarray}
\nonumber \m{Y}_1=\m{X}+\m{S}+\m{Z}_1\\
\m{Y}_2=\m{X}+\m{S}e^{j\phi}+\m{Z}_2
\end{eqnarray} 
For this model, the following upper bound holds:
\theorem  The rate for reliable communication for the channel model
described in Section~\ref{sec:prob} is bounded by 
\begin{equation}
R\leq \frac{1}{2}\log\left[\frac{(P+Q+1)^2}{4Q}\right]
\end{equation}
\proof
See Appendix~\ref{app:upperbd}.
\vspace{0.1in}

As an aside, we note that the bounding technique here can also be used
to find bounds on the rate for uncorrelated interference vectors of
\cite{khisti}. The bounds obtained are tighter than the bounds in
\cite{khisti} at high SIR.

\begin{figure}
\includegraphics[width=3.2in]{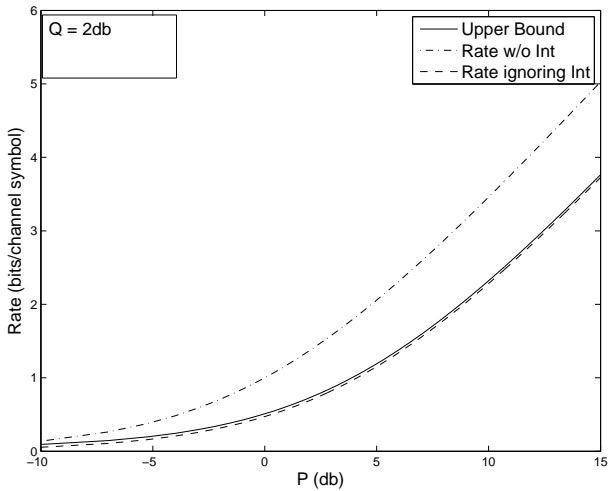}
\caption{Bounds on the rate for interference power $Q=2db$. The bound
  shows that at moderate SIR, ignoring interference and treating it as
  noise is close to optimal if we do not know the phase of the
  interference at the receiver.}
\label{fig:Q2db}
\end{figure}

\begin{figure}
\includegraphics[width=3.7in]{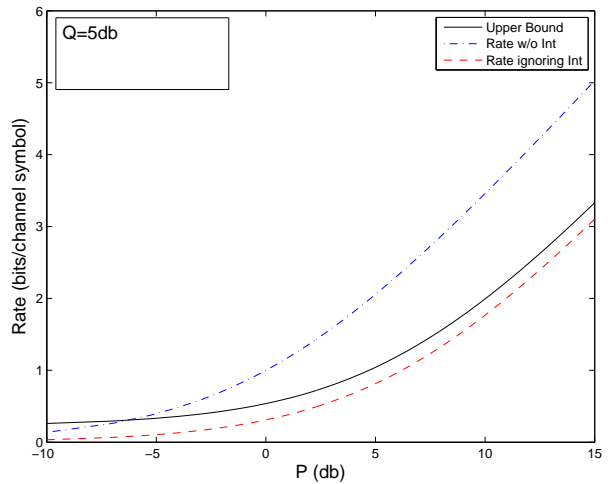}
\caption{Bounds on the rate for Q=5db. At low SIR, our new bound becomes
  very loose --- going above the rate with zero interference.}
\label{fig:Q5db}
\end{figure}

\begin{figure}
\includegraphics[width=3.7in]{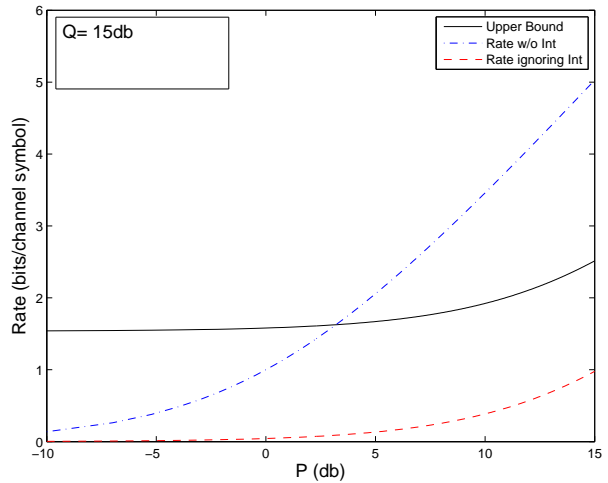}
\caption{The plot shows the bounds on the rate for $Q=15db$. Our upper
  bound is quite loose at low signal powers. Even at high signal
  powers, it is unclear if the bound is tight because we have no
  achievable schemes that approach this performance.}
\label{fig:Q15db}
\end{figure}

The bound for rate is plotted as a function of transmit power $P$ for
interference power $Q=2db$ in Fig. \ref{fig:Q2db}, for $Q=5db$ in
Fig.~\ref{fig:Q5db} and for $Q=15db$ in Fig.~\ref{fig:Q15db}. At low
$Q\approx 2db$, the bound is close to the rate achieved by ignoring
the interference knowledge completely and treating it as noise. This
shows that the bound is quite tight at low $Q$ and that missing a
single bit of phase-knowledge, makes the rest of the transmitter's
knowledge of the interference almost useless!  For increasing $Q$, the
bound is seen to get looser. In fact, it is evident from
Fig.~\ref{fig:Q15db} that the bound is extremely loose at low $P$ and
large $Q$ since it goes above the `no-interference' bound.

\begin{figure}
\includegraphics[width=3.5in]{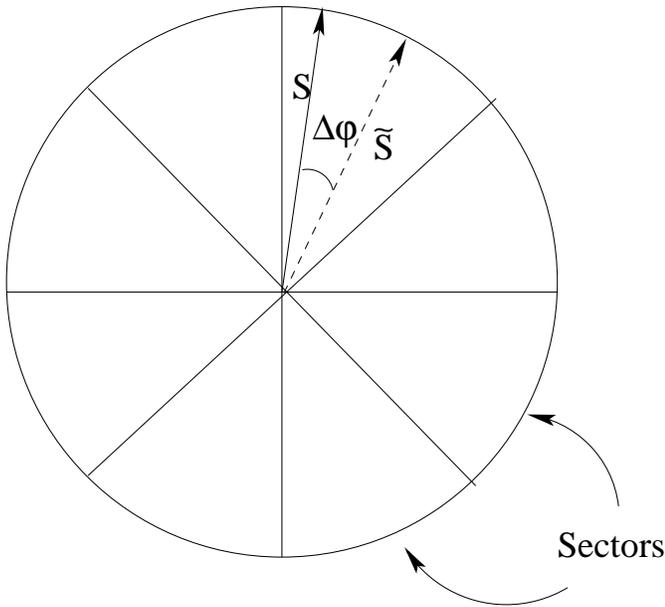}
\caption{The plot shows the idea behind the sectoring strategy. The phase-space is divided into $k=8$ sectors, and the transmitter dirty-paper codes according to the central vector in each sector, $\tilde{S}$. The actual interference is $S$, and the phase difference is $\Delta\phi$}
\label{fig:sectoringscheme}
\end{figure}

\section{Upper bound on the outage capacity under fade uncertainty of the interference}
\label{sec:bound}
In this section, we give an upper bound on the communication rate
assuming the fading coefficient of the interference is unknown at the transmitter.

Assume first that the fade amplitude $\gamma$ is fixed. We obtain computable upper bounds on the rate for given $\gamma$, using Theorem 1. Suppose the transmitter commits to a transmission rate $R$. We find the probability that the upper bound corresponding to the fade amplitude is below this rate. This gives a lower bound on the outage probability. 

We denote the capacity of the channel for given fixed $\gamma$ by $C(\gamma)$.

For this model, the following upper bound holds:
\corollary  The rate for reliable communication for the channel model
described in Section~\ref{sec:prob} with fixed fading coefficient $\gamma$ is bounded by\footnote{As an aside, we note that the bounding technique here can also be used
to find bounds on the rate for uncorrelated interference vectors of~\cite{khisti}. The bounds obtained are tighter than the bounds in~\cite{khisti} at high SIR.}

\begin{equation}
R\leq \frac{1}{2}\log\left[\frac{(P+\gamma^2Q+1)^2}{4\gamma^2Q}\right]
\end{equation}
The bound is denoted by $C_u(\gamma)$.\\
\proof
Follows from Theorem 1.\hfill{}$\Box{}$
\vspace{0.1in}

 \corollary Assuming that the transmitter now commits to a rate $R$, the outage probability is bounded by
\begin{equation}
p_{\mbox{out}} \geq \Pr(C_u(\gamma)<R)
\end{equation}
\proof 
The bound follows immediately from the fact that $p_{\mbox{out}}\geq \Pr(C(\gamma)<R)$ and $C(\gamma)\leq C_u(\gamma). \hfill{}\Box{}$.

The bound on outage probability depends on the fade distribution through the random variable $C_u(\gamma)$.

\begin{figure}
\includegraphics[width=4in]{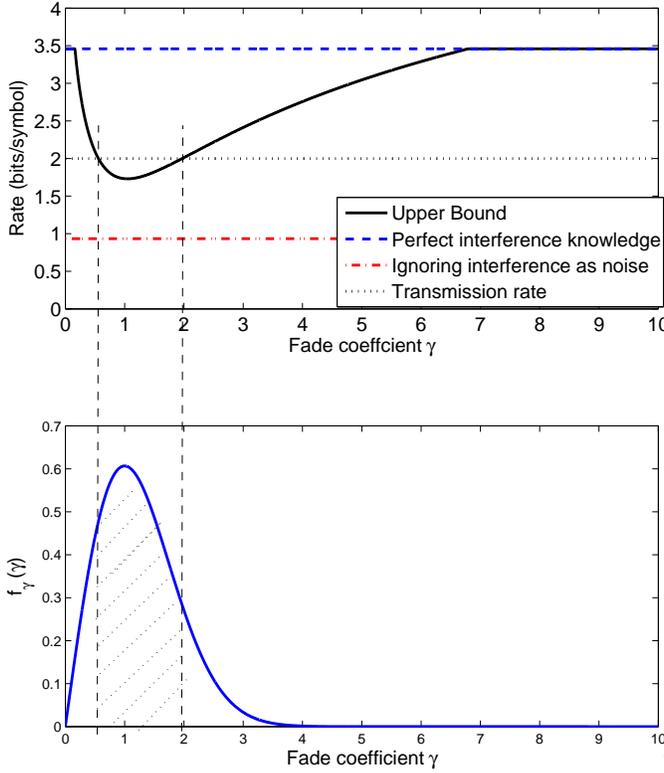}
\caption{The upper figure gives the upper bound on the rate for $P=Q=10db$ and varying fade coefficient $\gamma$. Also plotted are the rates with perfect knowledge of the interference (the classical dirty-paper coding) and the rate ignoring the knowledge of the interference.
The lower figure is the pdf of Rayleigh distribution for $\sigma^2=1$. Assuming the transmission rate to be $R=2$ bits/symbol, the calculation of lower bound on outage probability is shown. The lower bound is the area of the shaded region in the lower figure. The slowest increase in the rate with increase in outage probability would be when the peak of Rayleigh distribution lies near the valley of the bound. A surprising aspect of the upper bound is that for large $\gamma$, it is an increasing function of $\gamma$. We think this is an artifact of our bounding technique, and is not the actual behavior of the capacity.}
\label{fig:calculation}
\end{figure}

The bound on the rate is plotted as a function of the fade amplitude $\gamma$  for $P=Q=10\;db$ in Fig.~\ref{fig:calculation}. The lower figure in Fig.~\ref{fig:calculation} is the pdf of Rayleigh distribution for the parameter $\sigma^2=1$. For fixed rate $R=2$ bits/symbol, the probability $\Pr(R>C_u(\gamma))$ is the area of the shaded region in Fig.~\ref{fig:calculation}. This is a lower bound on the outage probability. Alternatively, for given outage probability, the curve gives an upper bound on the achievable rate. 

The bound on the rate in Fig.~\ref{fig:calculation} has a surprising feature: for large $\gamma$, it increases on increasing $\gamma$. We think this is an artifact of the bounding technique used, and is not fundamental to the problem at hand.

\begin{figure}
\includegraphics[width=4in]{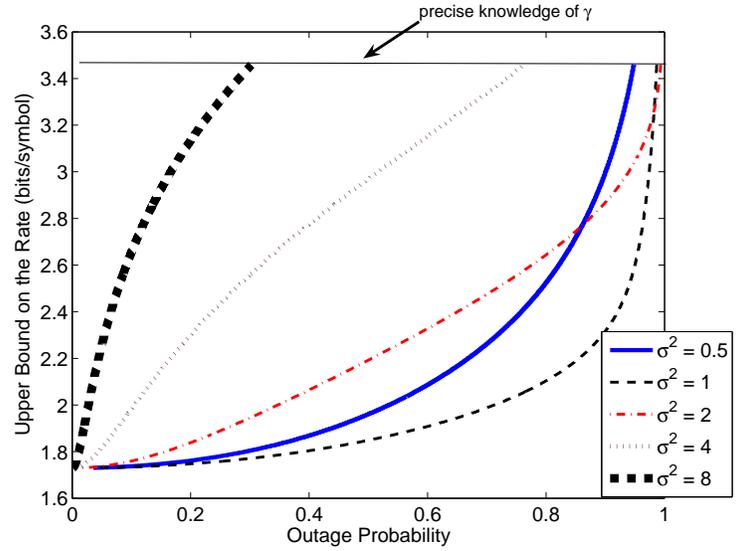}
\caption{The upper bound on the achievable rate vs the outage probability for varying parameter of Rayleigh distribution for $P=Q=10\;db$. From Fig.~\ref{fig:calculation}, it is evident that the slowest increase in outage probability would be when the mean of the Rayleigh distribution lies in the interval. This figure demonstrates the same explicitly. The capacity of channel with no uncertainty in $\gamma$ is $\sim 3.46$ bits/symbol. The outage probability is, therefore, 1 above this value. We think that the bound in Fig.~\ref{fig:calculation} is loose at high SIR. Therefore, the bounds for large values of $\sigma^2$ are also loose. Observe that all the curves converge to the same value for low outage probability. This value is the same as the minimum of the upper bound in Fig.~\ref{fig:calculation}. This is because the bound for outage probability of zero is a bound to the achievable rate for all values of $\gamma$.}
\label{fig:diffrayl}
\end{figure}

\begin{figure}
\includegraphics[width=4in]{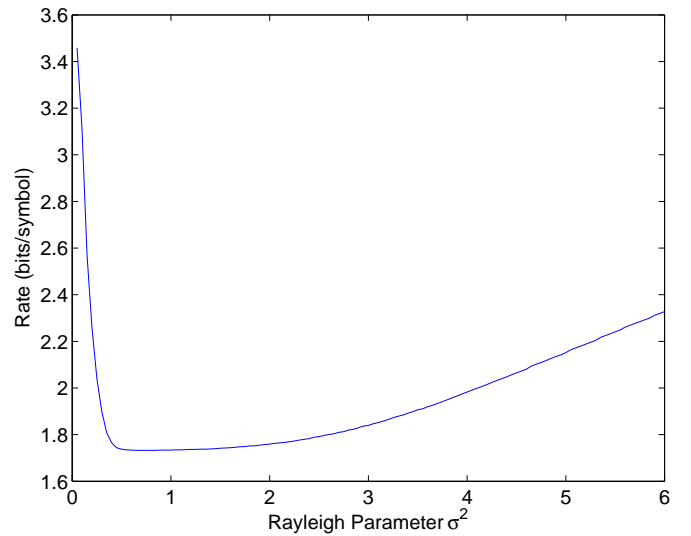}
\caption{The upper bound on the achievable rate vs the Rayleigh parameter $\sigma^2$ for $P=Q=10\;db$ and an outage probability of $0.1$. We think that the bound in Fig.~\ref{fig:calculation} is loose at high SIR. Therefore, the bound is loose at large $\sigma^2$. We think that the outage capacity is a decreasing function of $\sigma^2$.}
\label{fig:ratevsraylpara}
\end{figure}

The bound thus obtained is plotted as a function of outage probability for various Rayleigh parameters in Fig.~\ref{fig:diffrayl}, and as a function of Rayleigh parameter for outage probability of $0.1$ in Fig.~\ref{fig:ratevsraylpara}. Since the bound in Fig.~\ref{fig:calculation} is loose for large $\gamma$, the same is expected for the bound in Fig.~\ref{fig:ratevsraylpara} for large $\sigma^2$.

\section{A scheme for achieving improved rates without phase knowledge}
\label{sec:sector}
The bound above raises a natural question. Is it at all useful to have
the interference knowledge, without the phase? The curves in
Fig.~\ref{fig:Q5db} suggest that our bound is loose at low SIR,
therefore, it is likely that the greatest advantage lies there.

At low SIR, a natural strategy is to break the interference
uncertainty `circle' into a few sectors. Following \cite{khisti}, the
transmitter time shares between coding strategies that dirty paper
code according the `central' interference vector in each sector. We
call this strategy `sectoring'. The strategy is illustrated in \ref{fig:sectoringscheme}. To analyze such a strategy, however,
requires us to find the impact of small phase uncertainty on the
achievable rates. The problem is non-trivial because the residual
interference is not independent of the interference vector, and hence
not independent of the codeword to be detected.

\subsection{Achievable rates with small uncertainty in the phase}

Assuming that the uncertainty in phase is a small value $\Delta\phi$,
we investigate achievable rates for this reduced uncertainty. We wish
to dirty paper code with respect to the 'central' interference vector
in the uncertainty set, and treat the 'residual' uncertainty as noise.

Assume that the the phase $\phi\in [-\Delta\phi,\Delta\phi]$ for some
$0<\Delta\phi<\frac{\pi}{2}$. The transmitter chooses the codeword
$\m{U}$ according to the usual rule $\m{U}=\m{X}+\alpha\m{S}$, where
$\m{X}$ is chosen independent of $\m{S}$ as in \cite{costa}. The value
of $\alpha$ depends on $\Delta\phi$, as well as the SNR. The output
$\m{Y}$ is given by
\begin{eqnarray}
\m{Y}&=&\m{X}+\m{S}e^{j\phi}+\m{Z}\\
&=&\m{U}-\alpha\m{S}+\m{S}e^{j\phi}+\m{Z}
\end{eqnarray}
Since $\m{X}$ and $\m{S}$ are independent, $\m{U}$ is jointly Gaussian
with \m{S}. Therefore, $\m{U}$ and $\m{Y}$ are also jointly
Gaussian\footnote{Any linear combination of these two random variables
  is jointly Gaussian}.

We prove the following:
\theorem For a phase uncertainty of $\Delta\phi$, an achievable rate is given by\begin{equation}
R=\sup_{\alpha\in[0,1]}\log\left(\frac{P}{\mathcal{\bar{E}}(\Delta\phi)}\right)
\end{equation}
where 
\begin{eqnarray*}
\mathcal{\bar{E}}(\Delta\phi)=(1-\beta(\Delta\phi))^2P+(\alpha^2+\beta(\Delta\phi)^2
\\-2\alpha\beta(\Delta\phi) \cos(\phi))Q+\beta(\Delta\phi)^2
\end{eqnarray*}
and $\beta(\theta):=\frac{P+\alpha\cos(\theta)Q}{P+Q+1}$

\proof See Appendix~\ref{app:phaseuncertainty}.

\subsection{Rates achieved by sectoring}
Suppose we divide the circle of uncertainty into $k$ sectors, each
forming an angle of $\frac{2\pi}{k}$. Then we time-share dirty paper
code with respect to the central interference vector in each sector.
Intuitively, time sharing costs us a factors of $k$ in rate.  However,
for the best sectors, uncertainty in phase is at worst
$\frac{\pi}{k}$.

The rate achieved by sectoring is therefore given by
\begin{equation}
R=\frac{1}{k}\log\left(\frac{P}{\mathcal{\bar{E}}(\frac{\pi}{k})}   \right)
\end{equation}
To get an intuition into the achievable rates, we do some approximate
calculations. The residual uncertainty is (approximately)
$Q\;\sin^2(\pi/k)$. At low SIR,
\begin{equation}
R\approx \frac{1}{k}\log\left(1+\frac{P}{1+Q\sin^2(\pi/k)}\right)\approx \frac{1}{k}\frac{Pk^2}{Q\pi^2}
\end{equation}
if $Q\sin^2(\pi/k)>>1$. The optimal $k$ can be determined by
maximizing the rate over $k$. Intuitively, if $P<1$ and $Q>>1$, then
we should keep increasing number of sectors till $Q\pi^2/k^2\approx 1$
(residual interference becomes comparable to 1). For this value of
$k$, the rate obtained is $O(\left(\frac{P}{\sqrt{Q}}\right)$, which
is a gain of $\sqrt{Q}$. If $Q>>P>1$, then we should keep increasing
number of sectors till $Q\pi^2/k^2\approx P$ (the log approximation
fails at this point). The resulting rate is 
$O\left(\sqrt{\frac{P}{Q}}\right)$, a smaller advantage than for
$P<1$.

\begin{figure}
\includegraphics[width=3.7in]{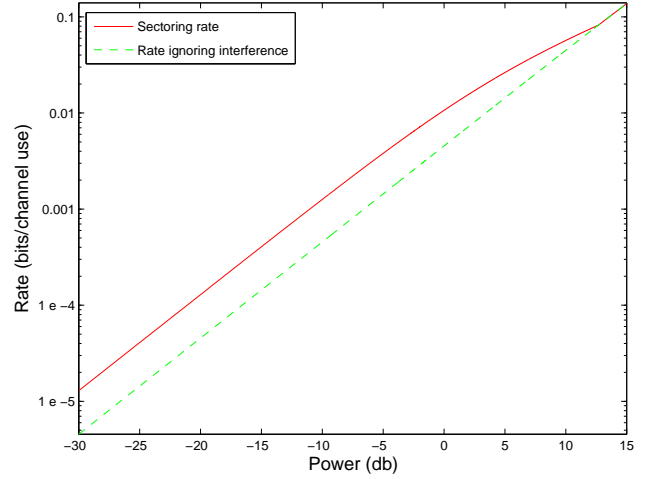}
\caption{Plotted on log scale, the difference between the two curves converges to a constant at low SIR. Thereby, the ratio of the two rates converges to a constant as SIR decreases.}
\label{fig:sector2}
\end{figure}

\begin{figure}
\includegraphics[width=3.7in]{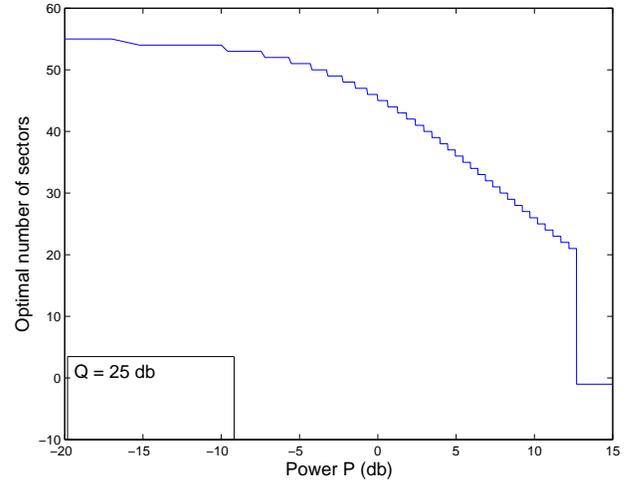}
\caption{Number of sectors required to achieve optimal performance decreases as SIR increases. 0 sectors corresponds to ignoring interference as noise. Q is again maintained at 25db.}
\label{fig:nos}
\end{figure}

We plot the (low SIR) region in which the sectoring method proposed
here performs better than ignoring interference as noise (see
Fig.~\ref{fig:sector2}). Fig.~\ref{fig:nos} shows the optimal number
of sectors with $P$ for fixed interference power of $25$ db. As
expected, an advantage is obtained only in extremely low SIR region.


\section{Performance with feedback: phase estimation}
\label{sec:feedback}
In Appendix~\ref{app:phaseuncertainty}, we derived a lower bound on
rate for some uncertainty in the phase. This can happen if the
transmitter gets some feedback from the receiver about phase estimate.
This avoids the need to waste energy on transmitting for all possible
phase sectors.

Traditionally, the problem of phase estimation is overcome by the
transmission of some pilot signal, which the receiver already knows.
However, it is not reasonable to assume that the primary transmitter
has a pilot signal known to the secondary receiver. Therefore, the
receiver cannot estimate the phase without the secondary transmitter's
help, since it does not know the signal transmitted by the interferer.

Since the interference phase uncertainty is zero-rate --- it does not
grow with time --- it is natural to ask if the receiver can estimate
the phase and convey it to the transmitter at an asymptotically
vanishing cost. In this section, we provide some ideas for phase
estimation, and give a scheme to combat interference in the face of
residual small uncertainty.

\subsection{Strategy for phase estimation}
Consider the following simple strategy to estimate phase. During time
instants $\{1,2,\ldots,k\}$ (call this the ``preamble'' in analogy
with training in traditional systems), the transmitter transfers its
`prescient knowledge' of the $S_k^{k+r-1}$ interference symbols by
appropriately compressing them and encoding the compressed data to
send over the channel. For this transmission, the receiver treats the
interference as Gaussian noise of power $Q$. Now, the receiver
estimates the phase during the first $r$ time instants, reducing the
uncertainty of phase to $\Delta\phi$, which depends on channel noise
as well as error introduced in quantization. The receiver then feeds
back the phase estimate to the transmitter.\footnote{The analogy may
  be more accurate with the RTS/CTS messages in many wireless
  protocols. The ``training packet'' plays the role of the RTS. The
  receiver responds with a small ``phase packet'' giving the
  interference phase to use --- in analogy with a CTS message. Only
  then can the main data transmission proceed. The difference is that
  here, this is a PHY-level consideration since it impacts waveform
  design rather than a MAC-level issue.} Finally, the transmitter
encodes the actual data payload for the remainder of the time. The
transmission ends at the end of a packet or the end of the coherence
time, whichever comes first.

The above strategy is well suited for a packet-oriented communication
scheme where the time between packets is potentially many channel
coherence times and unknown to the transmitter in advance. There is
not much we can do in such scenarios since the architecture dictates
that every packet must be self-sufficient. The price of this
self-sufficiency is that the ``preamble'' faces huge noise levels,
since the interference is unknown while decoding it. This imposes a
significant rate penalty, particularly under high SIR conditions.

The way around this problem is to assume a more continuously streaming
communication model in which data will be transmitted
block-after-block for many contiguous coherence times. In such cases,
the block-fading model here is questionable and a  Gauss-Markov model
is probably more appropriate. However, inspired by \cite{zheng}, we hope
that channel coherence issues are qualitatively the same in the two
cases and stick with our block-fading model. 

Two stages of transmission are introduced.  During the initialization
stage, a ``preamble'' as above is used. But once the transmitter
acquires the phase to a certain degree of accuracy, it
dirty-paper-codes the data {\em as well as} prescient knowledge of the
first $r$ symbols of the {\em next} coherence time period and sends
them to the receiver. In effect, the preamble's payload is combined
with the main data payload. This is possible because unlike a pilot
tone or a PRN-sequence used in a traditional preamble, we are not
interested in the waveform properties of the preamble
transmission. The relevant waveform for sounding out the interference
channel is provided by the primary transmission --- our only goal is
to empower the receiver to utilize it. 

\begin{figure}
\includegraphics[width=3in]{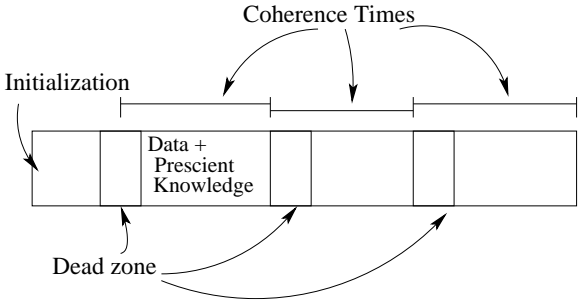}
\caption{Bootstrap strategy for phase estimation. The dead-zone 
  allows for phase estimation. Transmitter sends some prescient
  knowledge of the interference to the receiver during the previous
  data block. The receiver uses this knowledge to estimate phase
  during the dead-zone. The process is repeated in next coherence time.}
\label{fig:bootstrap}
\end{figure}

\begin{figure}
\begin{center}
\includegraphics[width=4in]{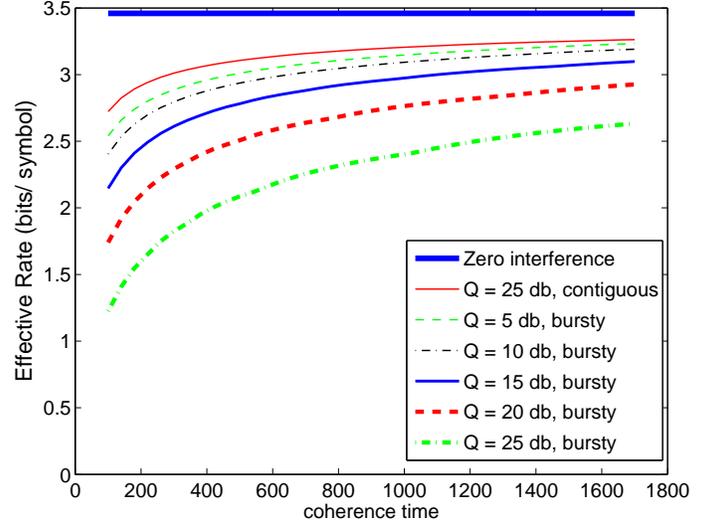}
\caption{Variation of effective rate achieved for $P=10db$ for bursty and non-bursty (contiguous time slots) transmissions. The bursty transmission results in significant loss of rate, with loss increasing for higher interference power(full phase knowledge
  Capacity$\sim3.46$ bits/symbol). For contiguous transmissions, increased coherence time allows for longer dead-zone with little loss in the rate. The variation of effective rate with $Q$, however, is not significant if we assume that the communication proceeds over contiguous coherent times since the data never really sees the interference. For the same relative fidelity of training quantization, a higher $Q$ allows for better phase estimation in the dead-zone, thereby reducing the effective interference indirectly in proportion to its increase. This assumes that the higher resolution feedback message does not significantly tax the feedback link capacity.} 
\label{fig:bursty}
\end{center}
\end{figure}

At the start of next coherence time, the decoder estimates the phase
using the first $r$ time instants and feeds this information back to
the transmitter. Once this has happened, the process repeats with the
next data block combined with the next preamble payload. The strategy
is figuratively illustrated in Fig.~\ref{fig:bootstrap}. Figure
\ref{fig:bursty} (top curve) demonstrates the performance of this 
strategy with coherence time for $Q=10db$. The underlying tension is between obtaining a good estimate
for the phase to improve data rate and wasting valuable time while
calculating that estimate. A higher value of interference allows for better estimation for the same dead-zone length. Therefore, the effective rate for increasing $Q$ does not show a significant change.

The strategy is detailed and analyzed in Appendix~\ref{app:strategy}. The analysis is approximate because we assume the residual phase error to be small. For small values of the interference power $Q$, the receiver may as well ignore interference altogether, rather than spending time learning it. In that case, the phase uncertainty is $\pi$, and the analysis is not valid. Therefore, the analysis holds only for large $Q$. The accuracy can be estimated from the phase value that attains the maximum rate.
 
Sometimes the transmitter may not know when its next transmission would be. This situation can occur when the source generates data at random times, unknown to the transmitter in advance. In this case, the initialization step needs to be performed for each packet, and therefore, the prescient knowledge has to be decoded treating all interference as noise. After decoding prescient knowledge, the transmitter keeps silent for the 'dead zone', as the receiver uses the prescient knowledge to estimate the phase. The receiver then feeds the phase back, and transmitter dirty-paper codes in accordance with the phase knowledge, as before.

This scheme is also analyzed at the end of Appendix~\ref{app:strategy}. The performance of this scheme is compared with that of the earlier strategy in Fig.~\ref{fig:bursty}.

\section{Conclusions and future work}
We showed that the dirty paper coding strategy requires the knowledge
of the interference phase at the transmitter. At high SIR, there is a
significant loss in performance due to not knowing the phase.  At
extremely low SIR, we provided a strategy that is able to utilize the
knowledge of the interferer's signal to \textit{some} extent without
having to get an estimate of the phase. We showed that if substantial
gains are desired, then it is worthwhile to engineer systems with
feedback where the transmitter also spends some of its power in
helping the receiver estimate the interference channel.

There are many open problems here:
\begin{itemize}
 \item The achievable schemes need to be tightened for both the case
       of no phase knowledge as well as when feedback is available.

 \item The sensitivity for multiple fading parameters beyond just
       phase needs to be established. Does every additional
       uncertainty impose its own overhead or does the case of phase
       uncertainty already capture all the essential effects?

 \item The upper bounds on the rate need to be tightened. The
       current bound is useless at low SINR and the current two-point
       uncertainty bound is also unable to adequately limit performance
       when the interferer is strong.

 \item The results need to be extended to Gauss-Markov models for
       fading where the variation is more continuous than the
       block-fading model here assumes.
\end{itemize}

In the end, the results in this paper are ``negative results'' that
suggest that attempts to exploit transmitter knowledge of the
interference signal realization will encounter engineering
difficulties. 


\appendices
\section{Proof of Theorem 1}
\label{app:upperbd}
Following \cite{khisti}, we assume that the transmitter has phase
uncertainty in the two point set $\{0,\pi\}$. That is, the
transmitter knows that the interference vectors is either $\m{S}$ or
$-\m{S}$. 
 
Remember that our model is
\begin{eqnarray}
\nonumber \m{Y}_1=\m{X}+\m{S}+\m{Z}_1\\
\m{Y}_2=\m{X}-\m{S}+\m{Z}_2
\end{eqnarray} 
Observe that the performance at any receiver is determined only by the
marginal distribution of the noise vectors at that receiver. We are
free to introduce \textit{any} joint distribution on the noise
vectors. For finding the tightest upper bound, we assume
$\m{Z}_1=\m{Z}_2$, that is, their correlation is 1.

In \cite{khisti}, the interference vector lies in a two-point set of
statistically independent vectors. In the bounding technique there,
one of the interference vectors is assumed to be zero. Using common
randomness between the receiver and the transmitter, they show that
this situation yields higher capacity than independent interferences
of the same power. We note that for our case, the method in
\cite{khisti} to upper bound the capacity does not work. This is
because the common randomness argument hinges on ability to generate
random sequences at the transmitter and the receiver that simulate
the interference (\cite{khisti}). For a phase shifted interference
vector, such sequence generation is not possible at the receiving end
since it does not know the interference realization. A simple way of
seeing this is where $\phi=0$. In this case, the usual dirty-paper
coding scheme is optimal. However, assuming one of the possible
interference vectors to be zero would decrease the rate beyond that! 

Since the transmitter is able to communicate reliably for either of
these two phases, the cut-set bound and Fano's inequality \cite{cover}
give us:
\begin{equation}
\label{eq:cutset}
nR\leq I(W;\m{Y}_i)+n\epsilon
\end{equation}
And therefore,
\begin{equation}
\label{eq:min2}
nR\leq \min_i I(W;\m{Y}_i)+n\epsilon
\end{equation}

To avoid clutter, we drop the $\epsilon$'s in the following steps. From \eqref{eq:min2},
\begin{eqnarray}
\label{eq:fano}
\nonumber nR\leq \frac{I(W;\m{Y}_1)+I(W;\m{Y}_2)}{2}\\
\nonumber = \frac{1}{2}(h(\m{Y}_1)-h(\m{Y}_1|W)+h(\m{Y}_2)-h(\m{Y}_2|W))\\
\leq \frac{1}{2}(h(\m{Y}_1)+h(\m{Y}_2)-h(\m{Y}_1,\m{Y_2}|W))
\end{eqnarray}
Suppose the correlation between $X_i$ and $S_{k,i}$ is $\rho_{ki}$ for
$k=1,2$. We observe that for random variables of given second moment,
the Gaussian random variable has the maximum entropy.  Thus,
\begin{equation}
\label{eq:yentropy}
h(\m{Y}_k)\leq \sum_{i=1}^n \log(2\pi e(P_i+Q+2\rho_{ki}\sqrt{P_iQ}+N))
\end{equation}
where $P_i$ is the power of the $i^{th}$ transmitted symbol, $X_i$. Also,
\begin{eqnarray}
\label{eq:y1y2w}
\nonumber
h(\m{Y}_1,\m{Y}_2|W)=h\left(\frac{\m{Y}_1+\m{Y}_2}{\sqrt{2}},\frac{\m{Y}_1-\m{Y}_2}{\sqrt{2}}\bigg{|}W\right)\\
=h\left(\frac{\m{Y}_1-\m{Y}_2}{\sqrt{2}}\bigg{|}W\right)+h\left(\frac{\m{Y}_1+\m{Y}_2}{\sqrt{2}}\bigg{|}W,\frac{\m{Y}_1-\m{Y}_2}{\sqrt{2}}\right)
\end{eqnarray}
Since $\m{Y}_1-\m{Y}_2$ is independent of $\m{X}$ (and hence, of $W$), the first term in \eqref{eq:y1y2w} is easy to simplify
\begin{eqnarray}
\label{eq:y1-y2}
\nonumber h\left(\frac{\m{Y}_1-\m{Y}_2}{\sqrt{2}}\bigg{|}W\right)&=&n\log(2\pi e Q(1-\cos\pi))
\\& = &n\log (4\pi e Q)
\end{eqnarray}
where we use $\m{Z}_1=\m{Z}_2$.

The second term can be lower bounded as follows
\begin{eqnarray}
\label{eq:y1+y2lower}
\nonumber
h\left(\frac{\m{Y}_1+\m{Y}_2}{\sqrt{2}}\bigg{|}W,\frac{\m{Y}_1-\m{Y}_2}{\sqrt{2}}\right)\\
\geq h\left(\frac{\m{Y}_1+\m{Y}_2}{\sqrt{2}}\bigg{|}W,\frac{\m{Y}_1-\m{Y}_2}{\sqrt{2}},\m{S},\m{X}\right)
\end{eqnarray}
Notice that $\m{Y}_1+\m{Y}_2=2\m{X}+2\m{Z}_1$. Therefore,
\begin{eqnarray}
\nonumber h\left(\frac{\m{Y}_1+\m{Y}_2}{\sqrt{2}}\bigg{|}W,\frac{\m{Y}_1-\m{Y}_2}{\sqrt{2}}\right)\geq h(\frac{2\m{Z}}{\sqrt{2}})
\\= n\log(2\pi e*2N)
\end{eqnarray}
Using \eqref{eq:y1y2w}, we can now lower bound the entropy
\begin{eqnarray}
\label{eq:lowbdy1y2w}
\nonumber h(\m{Y}_1,\m{Y}_2|W)\geq n\log(4\pi e Q)\\+n \log(4\pi e N)
\end{eqnarray}
Now in \eqref{eq:fano}, it is sufficient to bound $h(\m{Y}_1)+h(\m{Y}_2)$. Using \eqref{eq:yentropy},
\begin{eqnarray}
\label{eq:hy1+hy2}
\nonumber
&h(\m{Y}_1)+h(\m{Y}_2)
\\\nonumber& \leq \sum_{k=1,2}\sum_{i}\log[2\pi e(P_i+Q+2\rho_{ki}\sqrt{P_iQ}+N)]
\\\nonumber& = \sum_{i}\sum_{k=1,2}\log[2\pi e(P_i+Q+2\rho_{ki}\sqrt{P_iQ}+N)]
\\\nonumber& \overset{(a)}{\leq}  2\sum_{i}\log[2\pi e(P_i+Q+(\rho_{1i}+\rho_{2i})\sqrt{P_iQ}+N)]
\\\nonumber& \overset{(b)}{=}  2\sum_{i}\log[2\pi e(P_i+Q+N)]
\\\nonumber & \overset{(c)}{\leq} 2n \log[2\pi e(P+ Q+N)]
\end{eqnarray}
Inequalities $(a)$ and$(c)$ follow from concavity of $\log(\cdot)$ function. $(b)$ follows from the fact that $\rho_{1i}=-\rho_{2i}$. From \eqref{eq:lowbdy1y2w} and \eqref{eq:hy1+hy2}, we get the following bound on the rate:


\begin{equation}
R\leq \frac{1}{2}\log\left[\frac{(P+Q+N)^2}{4QN}\right]
\end{equation}
\begin{equation*}
\hspace{2.5in}\Box{}
\end{equation*}


\section{Derivation of lower bound for partial phase knowledge}
\label{app:phaseuncertainty}
\subsection{Analysis for $\phi=\Delta\phi$}
\label{sec:phiequaldeltaphi}

In the following, we use the techniques developed in \cite[Pg. 530--532]{davidbook} and \cite{coverchiang}.

We first detail the encoding and decoding scheme, and then analyse them for finding achievable rates.
 
Generate a Gaussian $U-$codebook, where each element of each codeword is drawn i.i.d. $\mathcal{CN}(0,P+\alpha^2Q)$. Generate $N_U=\left(\frac{P+\alpha^2Q}{\mathcal{\bar{E}}}\right)^{n(1-\epsilon)}$ such codewords to form the codebook. Distribute them randomly into $2^{nR}$ bins. At the encoder, given message $m\in\{1,\ldots,2^{\lfloor nR\rfloor}\}$ and $\m{S}$, find $\m{U}$ that is jointly typical with $\m{S}$. Send $\m{X}=\m{U}-\alpha\m{S}$.

At the decoder, find the Linear Least Square Estimate (LLSE) of $\m{U}$ given
$\m{Y}=\m{X}+\m{S}+\m{Z}$, assuming $\phi$ is known, and its value is $\Delta\phi$. Let the average LLSE error in each direction be denoted by $\mathcal{\bar{E}}$. Construct a sphere of radius $n(1+\delta)\mathcal{\bar{E}}$ ($\delta>0$, small) around $\m{\widehat{U}}$. Decode to the $U-$codeword in the sphere, if a unique such codeword exists. Else, declare a decoding error.

We now proceed to analyse the above scheme.

An encoding error would happen if for given $\m{S}$ and the message $m$, there is no jointly typical pair $(\m{U},\m{S})$ in bin $m$. 

The probability that a pair $(\m{U},\m{S})$ is jointly typical, where each of the elements is drawn independently, is greater than $(1-\epsilon)2^{-nI(U;S)-n\epsilon}$ for sufficiently large $n$. Therefore, the expected number of jointly typical codewords in a bin is greater than $(1-\epsilon)2^{n\epsilon}$ if
\begin{equation}
\label{eq:posexp}
N_u\times (1-\epsilon)2^{-nI(U;S)-n\epsilon}\times 2^{-nR}
\end{equation}
has a positive exponent. If this condition is satisfied, then as in \cite{coverchiang}, the probability of error converges to zero. 

The expression $I(U;S)$ simplifies to
\begin{eqnarray*}
I(U;S)&=&h(U)-h(U|S)\\
&=& h(U)-h(X)\\
&=& \log\left(\frac{P+\alpha^2Q}{P}\right)
\end{eqnarray*}

Therefore, for a positive exponent in \eqref{eq:posexp},
\begin{equation}
\label{eq:achrate}
R\leq \log\left(\frac{P}{\mathcal{\bar{E}}}\right)
\end{equation}

There are two potential sources of decoding error. The transmitted codeword may not lie in $S_U$, or there may be another codeword $\m{\tilde{U}}$ in $S_U$.

Let us first find the LLSE error $\mathcal{\bar{E}}$.
The LLSE estimate gives us:
\begin{eqnarray}
\widehat{U}_i&=&\frac{Re[E[U^*Y]]}{E|Y|^2}Y_i\\
&=& \frac{P+\alpha\; \cos(\Delta\phi) Q}{P+Q+1}Y_i=:\beta(\Delta\phi)Y_i
\end{eqnarray}
where the LLSE error is given by
\begin{eqnarray}
\nonumber\mathcal{\bar{E}}&=&E[|U_i-\widehat{U}_i|^2]
\\\nonumber &=& E[|X_i+\alpha S_i-\beta(X_i+S_ie^{j\Delta\phi}+Z_i)|^2]
\\\nonumber &=& E[| (1-\beta)X_i + (\alpha - \beta e^{j\Delta\phi})S_i - \beta Z_i  |^2]
\\\nonumber &=&(1-\beta)^2P+(\alpha^2+\beta^2 - 2\alpha\beta \cos(\Delta\phi))Q+\beta^2
\end{eqnarray}
The mean of the cross terms is zero, since by the dirty-paper-coding
construction, $X_i$ is approximately uncorrelated with $S_i$. In each of the real and the imaginary axes, the LSE is $\mathcal{\bar{E}}/2$.

By the weak law of large numbers, with high probability, the transmitted $U-$ codeword falls in the sphere $\mathcal{S}_U$ of radius $n(1+\epsilon)\sqrt{\mathcal{\bar{E}}}$ centered at $\m{\widehat{U}}$. To find the achievable rate, we now need to bound the probability of some other codeword $\m{\tilde{U}}$ falling in $\mathcal{S}_U$.

A Gaussian codebook of power $P+\alpha^2Q$ can be generated in an alternative way. Choose any radial direction randomly in the $n-$dimensional space. Then choose a point in this radial direction with the appropriate conditional distribution. This construction works because Gaussian distribution has uniform distribution over any radial direction. 

Another such distribution is uniform distribution over a sphere of radius $n(P+\alpha^2 Q)$ in $n-$dimensional. To generate a codebook with distribution uniform over the whole sphere, choose a radial direction randomly, and then choose a point in the radial direction with the appropriate conditional distribution\footnote{which is \textit{not} the uniform distribution in the radial direction}. 

As $n\rightarrow\infty$, both the distributions concentrate towards the edge of the sphere. For gaussian codebook this is because of the weak law of large numbers, whereas for uniform codebook, it is because the volume in large dimensions concentrates around the surface of the sphere. For $n$ large enough, conditioned on choosing the same radial direction, distance of a typical point on the radial direction chosen according to Gaussian codebook distribution from a typical point chosen according to uniform codebook distribution  can be made smaller than $n\epsilon$ for any $\epsilon>0$ for $n$ large enough. Effectively, codewords of both the codebooks can be made arbitrarily close to each other by choosing $n$ large enough.

Therefore, the decoder can safely assume that the codewords were chosen uniformly. It decodes to the uniform codebook codeword, which is close to the gaussian codeword for the same message. With that assumption, the probability that some other codeword $\m{\tilde{U}}$ falls in the sphere $S_U$ is 
\begin{equation}
\frac{Vol(\mathcal{S}_U)}{Vol(U-space)}=\left(\sqrt\frac{\frac{\mathcal{\bar{E}}}{2}}{\frac{P}{2}+\alpha^2\frac{Q}{2}}\right)^{2n}
\end{equation}
Using the union bound, the probability that any other codeword falls in $\mathcal{S}_U$ converges to zero as long as the number of $U-$codewords is no greater than $N_U=\left(\frac{P+\alpha^2Q}{\mathcal{\bar{E}}}\right)^{n(1-\epsilon)}$ for some $\epsilon>0$.

Observe that the proof is fairly general. In particular, the proof also works for jointly gaussian $X,S,Z$ with arbitrary correlations. The uniform codebook argument is brought in to use the volume arguments. The gaussian codebook is necessary for using typicality arguments.
\subsection{Analysis for $\phi<\Delta\phi$}
In absence of knowledge of $\phi$ at the receiver, we perform the same operations as for $\phi=\Delta\phi$. The mean squared error now is
\begin{eqnarray}
\nonumber \mathcal{\bar{E}}_\phi=E[|U_i-\widehat{U}_i|^2]
\\\nonumber=E[|X_i+\alpha S_i-\beta (X_i+S_i e^{j\phi}+Z_i)|^2]
\\\nonumber=E[(1-\beta(\Delta\phi))^2|X_i|^2
\\\nonumber+(\alpha^2+\beta(\Delta\phi)^2-2\alpha\beta(\Delta\phi) \cos(\phi))|S_i|^2
\\\nonumber+\beta^2+\text{cross terms}]
\\\nonumber=(1-\beta(\Delta\phi))^2P+(\alpha^2+\beta(\Delta\phi)^2
\\-2\alpha\beta(\Delta\phi) \cos(\phi))Q+\beta(\Delta\phi)^2
\end{eqnarray}
The mean of the cross terms is zero, since by construction, $X_i$ is approximately uncorrelated with $S_i$. 

Since $|\phi|<\Delta\phi$, the mean square error is only smaller.
Therefore, for $|\phi|<\Delta\phi$ and for $n$ large enough, the
nearest codeword only comes nearer, for the same noise realization. Therefore, the nearest codeword still lies in the sphere. Since rest of the codewords are randomly chosen, uniformly in the sphere, using the same argument as in \ref{sec:phiequaldeltaphi}, the probability of any another $U-$codeword falling into the sphere is small.

Therefore nearest neighbor decoding achieves the rate given by
\eqref{eq:achrate}. \hfill $\Box{}$

\section{Analysis of proposed phase estimation scheme}
\label{app:strategy}
Denote by $\m{\widehat{S}}$ the receiver's estimate of the
interference vector $\m{S}$. The transmitter source-codes $S$ before
sending it to the receiver. Denote the induced distortion by $D$.
Since it is obtained from the source code for $\m{S}$, it looks like
(see \cite[Pg. 479]{gallagerbook})
\begin{equation}
\m{\widehat{S}}=\gamma \m{S}+\m{\zeta}
\end{equation}
where $\m{\zeta}$ is Gaussian with variance $\frac{(Q-D)Q}{D}$ and the
scaling factor $\gamma=\frac{Q-D}{D}$.

For $k$ time instants, there is no transmission (dead-zone, $\tau$).
During this time, the receiver estimates the phase and then sends it
back to the transmitter. The feedback is modeled as instantaneous
since the phase can be encoded using a very small number of bits.

The receiver correlates the received vector with the signal it
receives. Let the correlation random variable at $i^{th}$ instant be
denoted by $T_i$. Then,
\begin{eqnarray}
T_i=(\gamma S_i+\zeta_i)^*(S_ie^{j\phi}+N_i)
\end{eqnarray}
where $N_i$ is the noise of unit variance. $T_i$ can be simplified into
\begin{eqnarray}
\label{eq:T}
\nonumber
T_i=(\gamma|S_i|^2e^{j\phi} )\\+(\gamma S_i^*N_i+ \zeta_i^*S_ie^{j\phi}+\zeta_i^*N_i)
\end{eqnarray}
It is clear from here that the sensitivity of the estimate is
independent of what the phase $\phi$ actually is. Therefore, without
loss of generality, we assume $\phi=0$. Denote the second term in
\eqref{eq:T} by $\eta_i$. Suppose the receiver estimates for $\tau$
time instants. Then the error phase error is approximately
$\frac{\sum_i \eta_i}{\tau\gamma Q}$.

The probability that this phase error is greater than $\Delta\phi$ is
\begin{equation}
\Pr\left(\frac{\sum_{i=1}^\tau \eta_i}{\tau\gamma Q}>\Delta\phi\right)=\Pr\left(\frac{\sum_{i=1}^\tau \eta_i}{\sqrt{\tau}\gamma Q}>\sqrt{\tau}\Delta\phi\right)
\end{equation}
Then, for large $\tau$, by the central limit theorem
\begin{equation}
\frac{\sum_{i=1}^\tau \eta_i}{\sqrt{\tau}\gamma Q} \sim \mathcal{N}\left(0,\frac{E[\eta_i^2]}{\gamma^2 Q^2}\right)
\end{equation}
where $E[\eta_i^2]=\gamma^2Q+Q\frac{Q-D}{D}+\frac{Q-D}{D}$. Therefore,
\begin{equation}
\Pr(\frac{\sum_{i=1}^\tau \eta_i}{\sqrt{\tau}\gamma Q}>\sqrt{\tau}\Delta\phi)\approx \mathcal{Q}\left( \frac{\sqrt{\tau}\gamma Q\Delta\phi}{\sqrt{E[\eta_i^2]}} \right)
\end{equation}
where $\mathcal{Q}$ is the familiar $\mathcal{Q}-$function. Set this
probability of outage to some fixed $\beta$ (say, $10^{-3}$) that
reflects the acceptable probability of outage. The training time to
reach $\Delta\phi$ error in phase with confidence of $1-\beta$ is
given by
\begin{equation}
\tau(\Delta\phi, D)=\frac{(\mathcal{Q}^{-1}(\beta))^2 E[\eta_i^2]}{\gamma^2Q^2\Delta\phi^2}
\end{equation}

The number of data bits communicated $N_{data}$ is
\begin{eqnarray*}
N_{\mbox{data}}=\max_{\tau,D}\bigg(\big(l_{\mbox{coh}}-\tau \big)C\left(P\frac{l_{\mbox{coh}}}{l_{\mbox{coh}}-\tau}\right)-\tau \log\left(\frac{Q}{D}\right)\bigg)
\end{eqnarray*}
where $l_{\mbox{coh}}$ is the coherence time, and $C(P)=\log(\frac{P}{\mathcal{\bar{E}}})$.
The effective rate is therefore
\begin{equation}
R_{\mbox{eff}} = \frac{N_{\mbox{data}}} {l_{\mbox{coh}}}
\end{equation}

\subsection{Performance with bursty transmissions}
In this model, the transmitter no longer communicates the prescient knowledge with data bits. Instead,  the initialization stage needs to be performed for each data packet, resulting in a greater decrease in the effective rate. 
During the initialization, the transmitter sends its prescient knowledge to the receiver with average power $P_1$ for time $\tau_{pk}$. The receiver ignores the interferences completely, treating it as noise, and decodes the prescient knowledge. 
After initialization, the transmitter is silent for the deadzone time, and the receiver estimates phase during this time, and feeds it back to the transmitter. Then the transmitter uses this phase estimate to dirty-paper code for the given interference, and transmits a signal of average power $P_1$. We do not allow for different transmit powers for transmitting the prescient knowledge and the actual data. $P_1$ is given by
\begin{equation}
P_1 = \frac{l_{\mbox{coh}}}{l_{\mbox{coh}}-\tau_{pk}-\tau_{\mbox{dead}}}
\end{equation}

The data bits are now communicated at the rate
\begin{equation}
C(P_1)=\log\left(\frac{P_1}{\mathcal{\bar{E_\phi}}}\right)
\end{equation}
We now optimize over the distortion $D$, $\tau_{\mbox{pk}}$ and the dead-zone $\tau_{\mbox{dead}}$ to obtain the largest possible number of bits communicated.
\begin{eqnarray*}
N_{\mbox{data}}=\max_{\tau_{pk},D}\bigg(  (l_{\mbox{coh}}-\tau_{\mbox{dead}}-\tau_{pk})\times C(P_1)  \bigg)
\end{eqnarray*}
And the effective rate is given by
\begin{equation}
R_{\mbox{eff}}=\frac{N_{\mbox{data}}}{l_{\mbox{coh}}}
\end{equation}

\bibliographystyle{unsrt}
\bibliography{IEEEabrv,dyspan}

\begin{thebibliography}{10}

\bibitem{davidbook}
David Tse and Pramod Viswanath.
\newblock {\em {Fundamentals of Wireless Communication}}.
\newblock Cambridge University Press, 2005.

\bibitem{telatar}
I~Telatar and DNC Tse.
\newblock Capacity and mutual information of wideband multipath fading
  channels.
\newblock {\em {IEEE} Trans. Inform. Theory}, 46(4), 2000.

\bibitem{zheng}
L~Zheng, DNC Tse, and M~Medard.
\newblock Channel coherence in the low snr regime.
\newblock {\em \textit{submitted to} IEEE Transactions on Information Theory},
  2005.

\bibitem{verduconstellations}
A.~Lozano, A.~M. Tulino, and Sergio Verdu.
\newblock Optimum power allocation for parallel gaussian channels with
  arbitrary input distributions.
\newblock {\em {IEEE} Trans. Inform. Theory}, 52(7):3033--3051, July 2006.

\bibitem{wigger}
A~Lapidoth, S~Shamai, and M~Wigger.
\newblock On the capacity of fading mimo broadcast channels with imperfect
  side-information.
\newblock {\em 43rd Annual Allerton Conference on Communication, Control, and
  Computing, Sept. 28-30}, 2005.

\bibitem{lapidothshamai}
A~Lapidoth and S~Shamai.
\newblock "fading channels: how perfect need "perfect side information" be?".
\newblock {\em {IEEE} Trans. Inform. Theory}, 48(5):1118--1134, May 2002.

\bibitem{jindal}
N~Jindal.
\newblock {MIMO Broadcast channels with Finite Rate Feeback}.
\newblock {\em {IEEE} Trans. Inform. Theory}, 52(11), November 2006.

\bibitem{tarokh}
N.~Devroye, P.~Mitran, and Vahid Tarokh.
\newblock Achievable rates in cognitive radio channels.
\newblock {\em {IEEE} Trans. Inform. Theory}, 52(5), 2006.

\bibitem{viswanath}
A.~Jovicic and Pramod Viswanath.
\newblock Cognitive radio: An information-theoretic perspective.
\newblock {\em \textit{submitted to} IEEE Transactions on Information Theory},
  2006.

\bibitem{wu}
Wei Wu, Sriram Vishwanath, and Ari Arapostathis.
\newblock {Capacity of Gaussian Weak Interference Channels with Degraded
  Message Sets}.
\newblock In {\em Conference on Information Sciences and Systems}, 2006.

\bibitem{gelfand}
S~Gelfand and MS~Pinsker.
\newblock Coding for channels with random parameters.
\newblock {\em Problems of Control and Information Theory}, 9(1):19--31, 1980.

\bibitem{costa}
M.~H.~M. Costa.
\newblock Writing on dirty paper.
\newblock {\em {IEEE} Trans. Inform. Theory}, 29(3), 1983.

\bibitem{khisti}
Ashish Khisti, Uri Erez, Amos Lapidoth, and Gregory Wornell.
\newblock Carbon copying onto dirty paper.
\newblock {\em Submitted to IEEE Transactions on Information Theory, available
  at http://arxiv.org/abs/cs.IT/0511095}, 2006.

\bibitem{mitran}
P~Mitran, N~Devroye, and V~Tarokh.
\newblock On compound channels with side information at the transmitter.
\newblock {\em {IEEE} Trans. Inform. Theory}, 52(4):1745--1755, April 2006.

\bibitem{wirelessnetworkcoding}
Sachin Katti, Hariharan Rahul, Wenjun Hu, Dina Katabi, Muriel Medard, and Jon
  Crowcroft.
\newblock Xors in the air: Practical wireless network coding.
\newblock In {\em Proceedings of 2006 ACM SigCOMM}, Pisa, Italy, 2006.

\bibitem{cover}
TM~Cover and JA~Thomas.
\newblock {\em { Elements of Information Theory}}.
\newblock WIley, 1991.

\bibitem{coverchiang}
TM~Cover and M~Chiang.
\newblock Duality between channel capacity and rate distortion with two-sided
  state information.
\newblock {\em IEEE Transactions on Information Theory, vol. 48, no. 6}, 2002.

\bibitem{gallagerbook}
RG~Gallager.
\newblock {\em Information Theory and Reliable Communication}.
\newblock Wiley, 1968.

\end{thebibliography}

\end{document}